\documentclass{revtex4}

\usepackage{amsmath,amssymb}
\usepackage[dvips]{graphicx}
\usepackage{bm}
\usepackage{color}

\begin{document}

\title{Majorana bound state of a Bogoliubov-de Gennes-Dirac Hamiltonian in arbitrary dimensions}

\author{Ken-Ichiro Imura$^1$, Takahiro Fukui$^2$, Takanori Fujiwara$^2$}

\affiliation{$^1$Department of Quantum Matter, AdSM, Hiroshima University, 739-8530, Japan\\
$^2$Department of Physics, Ibaraki University, Mito 310-8512, Japan
}

\begin{abstract}
We study a Majorana zero-energy state bound to a hedgehog-like point defect 
in a topological superconductor described by a Bogoliubov-de Gennes (BdG)-Dirac type effective
Hamiltonian.
We first give an explicit wave function of a Majorana state by solving the BdG equation directly,
from which an analytical index can be obtained.
Next, by calculating the corresponding topological index,
we show 
a precise equivalence
between both indices to confirm the index theorem.
Finally, we apply this observation 
to reexamine the role of another topological invariant, {\it i.e.}, the Chern number
associated with the Berry curvature proposed in the study of protected zero modes
along the lines of topological classification of insulators and superconductors.
We show that the Chern number is equivalent to the topological index,
implying that it indeed reflects the number of zero-energy states.
%
Our theoretical model belongs to the BDI class from the viewpoint of symmetry,
whereas the spatial dimension $d$ of the system is left arbitrary throughout
the paper.
\end{abstract}

\date{\today}

\maketitle


\section{Introduction}
Topological classification of matter dates back to reinterpretation
of the Kubo formula for the Hall conductance in terms of 
the first Chern number
associated with the Berry connection and curvature.
\cite{TKNN:82, Kohmoto:85}
The integer quantum Hall effect (IQHE) was thus given a status 
as a prototype of topologically distinguishable states of matter.
An extensive use of this idea;
reclassifying many-body states by means of topological invariants,
has been, on the other hand,
started much recently.
In the topological classification, 
matters are classified into a periodic table of ten universality classes,
including $\mathbb{Z}$ and $\mathbb{Z}_2$ type topological insulators and
superconductors.
\cite{SchnyderRFL:08, Kitaev:08,Zir96,AltZir97}
The range of its applicability
has been further extended beyond such ``bulk classifications'' ,
which involves only the Berry connection in momentum ($\bm k$-) space,
to include
various kinds of topological defects and associated protected zero-modes.
\cite{TeoKan10,TeoKane:10b}

With the periodic table in hands, we still need to establish how to relate this
topological classification to reality. In the case of IQHE we did not have to
ask this question, since a physical quantity, i.e., the Hall conductance
was directly related to a topological invariant.
The same type of reasoning is possible for the $\mathbb{Z}_2$ 
topological insulator, since it can be regarded as the time reversal counterpart 
($\mathbb{Z}_2$-version) of the IQHE;
as quantum {\it spin} Hall insulator.
\cite{KaneMele05a,KaneMele05b,XuMoore06,QWZ06,BerZha06,Roy09}
As for the protected zero-modes, similar topological invariant, i.e.,  
the Chern number has been proposed, \cite{TeoKan10,TeoKane:10b}
which involves the Berry connection 
in both $\bm k$- and $\bm x$- (real) space.
Still, a sufficient reasoning why such topological invariants
guarantee the existence of zero modes is lacking.
As the Kubo formula for the Hall conductance is to the IQHE, 
so which physical quantity is to the protected zero-modes?
We are still in search for its precise analogue.

Specific two models, Dirac fermions with a monopole in 3D \cite{JackiwRebbi:76} as well as
those with a vortex in 2D\cite{JackiwRossi:81} have been known as prototypes of fermions with
protected zero-energy states. 
For these models, zero-energy wave functions have been 
explicitly obtained,\cite{JackiwRebbi:76,JackiwRossi:81}
and index theorem has been successfully established.\cite{Callias:78,Weinberg:81,FukuiFujiwara:10} 
Here, the index theorem \cite{Vol03} claims the equivalence between the analytical index 
concerning the zero-energy states and the topological invariant associated with the
configuration of the order parameter.
The Chern number with respect to the Berry curvature has also been calculated for these models.
\cite{TeoKan10,TeoKane:10b}
Remarkably, the Chern number and the topological index reached
the same expression, although the starting points are quite different.

For a closer inspection on
the proposed Chern number and its precise agreement with the topological index,
we study in this paper
a Majorana zero-energy state bound to a topological defect
in a BdG-Dirac type superconductor 
in $d$-spatial dimensions.
We first give an explicit wave function of the zero-energy state by solving the BdG equation directly.
Next, by calculating the corresponding topological index,
we confirm the index theorem. 
Finally, we apply this observation 
to reexamine the role of the Chern number
associated with the Berry curvature.
We show that the Chern number is equivalent to the topological index in arbitrary dimensions,
implying that it indeed reflects the number of protected zero-energy states.

%


The structure of the paper is as follows:
In Sec. \ref{s:Zer},  we construct normalizable zero-energy wave functions
of the model, and identify its analytic indices.
The equivalence of analytic and topological indices
is explicitly established, i.e.,
the index theorem is concretized in Sec. \ref{s:Ind}.
In Sec. \ref{s:Che} we reexamine the structure of topological
arguments based on the Berry connection and curvature,
and contrast it with those of earlier sections.
We establish an explicit one-to-one correspondence between
the index of the Hamiltonian and the topological Chern number.
In Sec. \ref{s:Sum}, summary and discussions are given.

\section{Zero-energy state of the BdG-Dirac Hamiltonian}
\label{s:Zer}

Let us consider a superconducting condensate described by the following
BdG-Dirac type Hamiltonian,
\begin{eqnarray}
{\cal H}=-i\Gamma^j\partial_j+\Gamma^{d+a}\phi_a ,
\label{MinMod}
\end{eqnarray} 
where $j,a=1,\cdots,d$.
In order to make our parallelism illustrative and convincing
The spatial dimension $d$ is kept arbitrary throughout the paper.
The $\Gamma$-matrices obey $\{\Gamma^\mu,\Gamma^\nu\}=2\delta^{\mu\nu}$
for $\mu$, $\nu=1,\cdots,2d$. We also define for later use a generalized chiral matrix by
$\Gamma^{2d+1}=(-i)^d\Gamma^1\cdots\gamma^{2d}$, which anti-commutes with
$\Gamma^\mu$.
The field $\phi_a$ denotes the $a$-th component of a given multicomponent order parameter.
In Eq. (\ref{MinMod})
two types of $\Gamma$-matrices, associated with the kinetic and the order parameter part
of the Hamiltonian,
have been introduced in the Nambu notation
for a system of Dirac fermions in an arbitrary spatial dimension $d$.
When $d=2$ the order parameter $\bm\phi$ has two components,
$\bm\phi = (\phi_1, \phi_2)$,
each corresponding to the real and the imaginary part of a gap function.
When $d=3$ the $\bm\phi$-field acquires a third component $\phi_3$,
which represents a mass term encoding a band inversion.
Such a 3-component order parameter has been proposed for describing proximity effects in
a topological insulator/superconductor hybrid system.
\cite{FuKane:08,TeoKan10}
The $d$-component order parameter field $\bm\phi$ in Eq. (\ref{MinMod})
is a natural generalization of this to the case of an arbitrary spatial dimension $d$.
In what follows we assume that the chemical potential is set, unless otherwise mentioned, to zero;
the model has chiral symmetry (see below for details).

Let us consider a point defect of $\bm\phi$ located at the origin, i.e.,
we assume that $\bm\phi$ takes the following hedgehog configuration,
\begin{eqnarray}
\phi_j(x)=\Delta(r)\hat x_j,
\label{Hig}
\end{eqnarray}
where $r$ is the radial coordinate and  $\hat x_j=x_j/r$. 
We assume $\Delta(\infty)=\mbox{const.}$ and impose $\Delta(0)=0$ 
to ensure the regularity of $\bm\phi$ at the origin. 
Physically, such a point defect may be associated with the presence of
a vortex or a monopole 
but here we assume that the hedgehog configuration (\ref{Hig})
is given as a background field, 
and do not take account of the effects of associated electromagnetic field.

The same (mathematical) model, Eq. (\ref{MinMod}), in cases of $d=2$ and $d=3$,
has been
extensively investigated, but in a different physical context.
\cite{JackiwRebbi:76, JackiwRossi:81,Callias:78, Weinberg:81}
%
In contrast to these existing studies,
the system we have in mind is a topological insulator 
under the proximity effect of a superconductor.
\cite{FuKane:08,TeoKan10}
Let us also mention that a variant of the same model with a specific choice of $\Delta(r)=r$ has been 
recently investigated in Ref. \cite{HerLu11}.
Compared with this model, our model assumes a realistic behavior of the gap function,
$\Delta(\infty)=$const., 
which makes the problem nontrivial even for zero-energy states. 


In the classification of topological insulators and superconductors, 
our setup, i.e., Eq. (\ref{MinMod}) with (\ref{Hig}),
belongs to the BDI symmetry class, 
\cite{TeoKane:10b}
which has both particle-hole and chiral symmetries.
Particle-hole symmetry is described by $\{{\cal H},C\}=0$, where $C$ is an antiunitary 
operator with $C^2=1$. With an appropriate choice of the $\Gamma$-matrices, 
it can be made $C=K$, where $K$ is the complex conjugation operator,
implying that the Hamiltonian is antisymmetric. \cite{CJNPS10}
The chiral symmetry of the model is described 
by $\{{\cal H},\Gamma^{2d+1}\}=0$. 
It may seem awkward to some readers,
since in conventional relativistic massless Dirac systems, 
chiral symmetry exists only in even spatial dimensions. 
In the present case, however, 
since chiral symmetry is extended to 
include the internal space of the order parameter field;
dimension is doubled,
chiral symmetry is allowed in arbitrary dimensions $d$.

We derive
normalizable zero-energy wave functions
describing electronic states bound to the defect. 
We deal with the case of $d$: odd and even separately,
due to the representation of the $\Gamma$-matrices.

\subsection{Zero-energy state: case of $d=2n+1$}
\label{s:OddDim}

To solve the eigenvalue equation for ${\cal H}$ it is convenient to 
work with a concrete representation of the $\Gamma$-matrices. 
In odd spatial dimensions, $d=2n+1$, the $\Gamma$-matrices can be chosen 
as 
\begin{eqnarray}
\label{gamma_1}
\Gamma^j=\gamma^j\otimes 1\otimes \sigma^1,\quad
\Gamma^{d+j}=1\otimes\gamma^j\otimes \sigma^2, 
\end{eqnarray}
where $j=1,\cdots,d$, and
\begin{eqnarray}
\label{gamma_1b}
\Gamma^{2d+1}=1\otimes1\otimes\sigma^3.
\end{eqnarray}
Here, $\gamma$'s are conventional $\gamma$-matrices given in Appendix \ref{s:Gam}.
This choice of the $\Gamma$-matrices have the following advantage;
the Hamiltonian matrix (\ref{MinMod}) 
decomposes into independent blocks of two chiral components,
\begin{eqnarray}
{\cal H}=
-i\left(
\begin{array}{cc}
&\gamma^j\otimes1\partial_j+1\otimes\gamma^j\phi_j\\
\gamma^j\otimes1\partial_j-1\otimes\gamma^j\phi_j&
\end{array}
\right) .
\end{eqnarray}
Correspondingly, let $\Psi=(\Psi_+,\Psi_-)^T$ be a chirally-decomposed zero-energy wave function;
when $\Psi_-=0$ ($\Psi_+=0$), the chirality of the wave function $\Psi$ is $+1$ ($-1$).
The eigenvalue equation, ${\cal H}\Psi=0$, can be written as
\begin{eqnarray}
\left(\gamma^j\otimes1\partial_j\pm1\otimes\gamma^j\phi_j\right)\Psi_\pm=0.
\label{OddZerEqu}
\end{eqnarray}
To solve the above equation, it is convenient to use the following notation:
\cite{JackiwRebbi:76}
the action of the matrix $a\otimes b$ to the wave function $\Psi$ is defined as
$(a\otimes b\Psi)_{\alpha\lambda}=
a_{\alpha\beta}b_{\lambda\rho}\Psi_{\beta\rho}=a_{\alpha\beta}\Psi_{\beta\rho}(b^T)_{\rho\lambda}
=(a\Psi b^T)_{\alpha\lambda}$, which enables us to regard $\Psi$ as a matrix-valued
wave function.
By the use of this notation, Eq. (\ref{OddZerEqu}) can be written as
\begin{eqnarray}
\gamma^j\partial_j\Psi_\pm\mp\Psi_\pm(\gamma^j)^T\phi_j=0.
\end{eqnarray}
If we note Eq. (\ref{ReaGam}), we have
\begin{eqnarray}
\gamma^j\partial_j\widetilde\Psi_\pm\mp(-)^n\widetilde\Psi_\pm\gamma^j\phi_j=0,
\end{eqnarray}
where we have introduced
$\widetilde\Psi_\pm=\Psi_\pm\Gamma$,
using $\Gamma$, a product of $\gamma$-matrices
defined as in (\ref{gamma_cap}), and 
$n$ is an integer specifying the spatial dimension $d=2n+1$.
For the hedgehog configuration (\ref{Hig}),
we may assume that $\widetilde \Psi_\pm$ depends only on $r$, namely,
$\partial_j\widetilde\Psi_\pm(r)=\hat x_j\widetilde\Psi'(r)$, where 
$\widetilde\Psi'(r)\equiv\partial_r\widetilde\Psi_\pm(r)$, and also that
the matrix wave function $\widetilde\Psi_\pm$ is diagonal, 
$(\widetilde\Psi_\pm)_{\alpha\beta}(r)=\delta_{\alpha\beta}\psi_\pm(r)$. 
It then follows that
\begin{eqnarray}
\psi'_\pm(r)\mp(-)^n\Delta(r)\psi_\pm(r)=0,
\end{eqnarray}
and its solutions given by
\begin{eqnarray}
\psi_\pm=C_\pm\exp\left[\pm(-)^n\int_0^r\Delta(r')dr'\right] ,
\end{eqnarray}
where $C_\pm$ is a numerical constant. 
Obviously, only either of $\psi_+$ or $\psi_-$ is normalizable depending on $n$ and
on the sign of $\Delta(\infty)$. 
Thus, the normalizable
zero-energy wave function is given by 
\begin{eqnarray}
\Psi=
\left(\begin{array}{c}\Psi_+\\0\end{array}\right)=
C_+\left(\begin{array}{c}\Gamma\\0\end{array}\right)
\exp\left[(-)^n\int_0^r\Delta(r')dr'\right] 
\label{psi+}
\end{eqnarray}
for $(-)^n\Delta(\infty)<0$, and
\begin{eqnarray}
\Psi=
\left(\begin{array}{c}0\\\Psi_-\end{array}\right)=
C_-\left(\begin{array}{c}0\\\Gamma\end{array}\right)
\exp\left[-(-)^n\int_0^r\Delta(r')dr'\right] 
\label{psi-}
\end{eqnarray}
for $(-)^n\Delta(\infty)>0$.
Eqs. (\ref{psi+}, \ref{psi-}) are indeed
chirally decomposed zero-energy wave functions,
i.e., the eigenvalue of Eq. (\ref{gamma_1b}) is
$+1$ for Eq. (\ref{psi+}),
whereas $-1$ for Eq. (\ref{psi-}).
Namely, they signify that for a given $\Delta(\infty)>0$
the chirality of the zero-energy state is $(-)^{n+1}$.
If $\Delta(\infty)<0$, on the other hand,
it is given by $(-)^n$.

\subsection{Zero-energy state: case of $d=2n$}
\label{s:EveDim}

The procedure of constructing the zero-energy wave function is similar.
The difference stems from the representation of the $\Gamma$-matrices:
In the present case, there is not such a convenient chirally-decomposed representation
as has been adopted in  odd dimensions. 
So that 
in even spatial dimensions, we use the following $\Gamma$-matrices 
\begin{eqnarray}
\label{gamma_2}
\Gamma^j=\gamma^j\otimes1,\quad \Gamma^{j+d}=\gamma^{2n+1}\otimes\gamma^j,
\end{eqnarray}
where $j=1,\cdots,d~(=2n)$, and
\begin{eqnarray}
\label{gamma_2b}
\Gamma^{2d+1} 
=\gamma^{2n+1}\otimes\gamma^{2n+1}.
\end{eqnarray}
Here, $\gamma$-matrices are also defined in Appendix \ref{s:Gam}.
The zero-energy wave function obeys
\begin{eqnarray}
-i\gamma^j\otimes1\partial_j\Psi+\gamma^{2n+1}\otimes\gamma^j\phi_j\Psi=0 .
\end{eqnarray}
By the use of the similar notation for the matrix-valued wave function 
in the previous subsection, we have
\begin{eqnarray}
-i\gamma^j\partial_j\widetilde\Psi+(-)^n\gamma^{2n+1}\widetilde\Psi\gamma^j\phi_j=0,
\quad \widetilde\Psi\equiv\Psi \Gamma.
\end{eqnarray}
Recall that $\Gamma$ is defined in (\ref{gamma_cap}).
Taking Eq. (\ref{Hig}) into account and 
assuming 
\begin{eqnarray}
\widetilde\Psi(r)=\psi(r)+i\gamma^{2n+1}\varphi(r) ,
\label{EveAns}
\end{eqnarray}
the l.h.s. of the 
above equation can be converted into 
\begin{eqnarray}
&-i\gamma^j(\psi'+i\gamma^{2n+1}\varphi')
+(-)^n\gamma^{2n+1}\Delta(\psi+i\gamma^{2n+1}\varphi)\gamma^j
\nonumber \\
&=-i\gamma^j\left[\psi'-(-)^n\Delta\varphi\right]
+\gamma^j\gamma^{2n+1}\left[\varphi'-(-)^n\Delta\psi\right],
\end{eqnarray}
where $\psi'=\partial_r\psi$ and $\varphi'=\partial_r\varphi$. 
This leads to the following coupled equations
\begin{eqnarray}
\psi'(r) &=& (-)^n\Delta(r)\varphi(r),\quad
\nonumber\\
\varphi'(r) &=& (-)^n\Delta(r)\psi(r) .
\end{eqnarray}
These equations formally allow two independent solutions
\begin{eqnarray}
\psi_\pm(r)=\pm(-)^n\varphi_\pm(r)=C_\pm\exp\left[\pm\int_0^r\Delta(r')dr'\right] .
\label{psvph}
\end{eqnarray}
The normalizability of the wave functions imposes   
$C_\pm=0$ for $\Delta(\infty)\gtrless0$. 
These two kinds of wavefunctions are reminiscent of those in 
Eqs. (\ref{psi+}) and (\ref{psi-}).
However, in either case of Eq. (\ref{psvph}), 
the chirality of the wave function is
\begin{eqnarray}
\Gamma^{2d+1}\Psi &=& \gamma^{2n+1}\Psi(\gamma^{2n+1})^T
\nonumber\\
&=& \gamma^{2n+1}\Gamma^\dagger\widetilde\Psi\gamma^{2n+1}
\nonumber\\
&=& (-)^n\Psi ,
\end{eqnarray}
where we have used the facts that 
$\widetilde\Psi\gamma^{2n+1}=\gamma^{2n+1}\widetilde\Psi$ for $\widetilde\Psi$ in 
Eq. (\ref{EveAns}) and $\Gamma^\dagger\gamma^{2n+1}=(-)^n\gamma^{2n+1}\Gamma^\dagger$
which follows from Eq. (\ref{ReaGam}). We thus find that the 
chirality of the zero-mode in $d=2n$ dimensions is given by 
$(-)^n$ independently of the sign of $\Delta(\infty)$. This 
is in contrast with the odd dimensional case described in the 
previous subsection: In even dimensions, $\bm x\rightarrow-\bm x$ 
does not cause space reflection and two configurations $\bm\phi$ and 
$-\bm\phi$ belong to the same topological sector. To obtain 
the zero-energy state with opposite chirality we may consider the following configuration;
\begin{eqnarray}
  \phi_j(x)=\Delta(r)\times
\left\{
\begin{array}{ll}
    \hat x_j & (j=2,3,\cdots,d) \\
    -\hat x_1 & (j=1)
\end{array}
\right.
\end{eqnarray}
This is a mirror reflection of Eq. (\ref{Hig}) with respect 
to the $x_1$-axis. It is easy to see that zero-energy wave function 
can be written as  
\begin{eqnarray}
  \Psi=\gamma^1(\psi(r)+i\gamma^{2n+1}\varphi(r)),
\end{eqnarray}
where $\psi$ and $\varphi$ are given by Eq. (\ref{psvph}). Due 
to the presence of $\gamma^1$ the chirality of the 
wave function becomes $-(-)^n$. 

\subsection{Index of the Hamiltonian: the analytic index}

As in the case of $d=2$ and $3$, we define the index of the Hamiltonian
with respect to its zero-energy state. 
Let $N_\pm$ be the number of zero-energy states of 
${\cal H}$ whose chirality, i.e., the eigenvalue of $\Gamma^{2d+1}$,
is $\pm1$, respectively.
Then the index of ${\cal H}$ is defined by
\begin{eqnarray}
{\rm ind}\,{\cal H} \equiv N_+-N_- .
\label{AnaInd}
\end{eqnarray}
Let us assume that the Hamiltonian Eq. (\ref{MinMod}) allows only one zero-energy
bound state, which has been obtained so far in Sec. \ref{s:OddDim} and \ref{s:EveDim}. 
Then,  provided that $\Delta(\infty)>0$, the index (\ref{AnaInd}) reads
\begin{eqnarray}
{\rm ind}\,{\cal H}=
\left\{
\begin{array}{ll}
(-)^n \quad & (d=2n)\\
(-)^{n+1} \quad & (d=2n+1)
\end{array}
\right. .
\label{AnaIndRes}
\end{eqnarray}

\section{Index theorem}
\label{s:Ind}

In the last section we have explicitly constructed chirally decomposed  
zero-energy states
in order to introduce and determine the {\it analytical} index of the Hamiltonian.
The obtained index is given in Eq. (\ref{AnaIndRes}) as a function of the
spatial dimension $d$.
These worked out thanks to specific representations of the 
$\Gamma$-matrices, e.g., Eqs. (\ref{gamma_1}) and (\ref{gamma_1b}), 
or Eqs.  (\ref{gamma_2}) and (\ref{gamma_2b}).
Here, we consider {\it a priori} another index of the Hamiltonian, 
called a {\it topological} index,
the evaluation of which
involves no specific representation of the $\Gamma$-matrices. 
The equivalence of the two indices is then guaranteed by 
the index theorem:
\cite{Callias:78,Weinberg:81,FukuiFujiwara:10}
\begin{eqnarray}
{\rm ind}\,{\cal H}
&=& \lim_{m\rightarrow0}{\rm Tr}\,\Gamma^{2d+1}\frac{m^2}{{\cal H}^2+m^2}
\nonumber\\
&=& c+
\lim_{m\rightarrow0}\frac{1}{2}\int_{|x|\rightarrow\infty} J^j(x) {\rm d} S_j.
\label{IndThe}
\end{eqnarray}
The first (second) line corresponds to analytical (topological)
representation of the index.
In the second line, $c$ is defined by
\begin{eqnarray}
c=\lim_{M\rightarrow\infty}{\rm Tr}\,\Gamma^{2d+1}\frac{M^2}{{\cal H}^2+M^2},
\label{CheMag}
\end{eqnarray}
and the current density $J^j(x)$ is generally expressed as
\begin{eqnarray}
J^j(x)&\equiv&
\lim_{y\rightarrow x}{\rm tr}\,\Gamma^{2d+1}\Gamma^j
\left(\frac{1}{-i{\cal H}+m}-\frac{1}{-i{\cal H}+M}\right)
\delta^d(x-y)
\nonumber\\
&=& \lim_{y\rightarrow x}{\rm tr}\,\Gamma^{2d+1}\Gamma^j
\left(\frac{i{\cal H}}{{\cal H}^2+m^2}-\frac{i{\cal H}}{{\cal H}^2+M^2}\right) 
\delta^d(x-y) .
\label{AxiVecCur}
\end{eqnarray}
where, we have introduced the Pauli-Villars regulator with mass $M$
to make the current well-defined, which yields the term $c$ in Eqs. 
(\ref{IndThe}) and (\ref{CheMag}).
\cite{Weinberg:81} 
In the second line of Eq. (\ref{IndThe})
the integration is taken over S$^{d-1}$, i.e., the boundary of R$^d$.
The first line of Eq. (\ref{IndThe}), or
the definition of index in Eq. (\ref{AnaInd})
reflects analytical properties of the Hamiltonian, as the solution of
the BdG equation as investigated in the previous section.
The second line of Eq. (\ref{IndThe}), on the contrary,
is associated with topological properties of the order parameter.

The Pauli-Villars regulator introduced in Eq. (\ref{AnaInd})
becomes particularly important,
when the gauge potential is taken into account;
the second term of (\ref{AxiVecCur}) gives, indeed, the Chern number 
associated with the magnetic field.
Here, on the contrary,
since the gauge potential is not considered,
the integration of the second term vanishes after $M\rightarrow\infty$.
The current density 
is indeed well-defined
even without the Pauli-Villars regulator. (Below, we will verify this explicitly.)
In the plane wave basis,
it can be calculated as
\begin{eqnarray}
J^j(x)=\int\frac{{\rm d}^dk}{(2\pi)^d}e^{-ikx}{\rm tr}\,\Gamma^{2d+1}\Gamma^j(i{\cal H})
\frac{1}{{\cal H}^2+m^2}e^{ikx}.
\label{CurMom}
\end{eqnarray}
Here, we have introduced an abbreviation for the inner product, $kx=k_jx_j$.
In the following, 
we evaluate this expression,
i.e., carrying out the integration over the momentum in Eq. (\ref{CurMom})
in two different ways.

\subsection{Direct calculation of the topological index}
\label{s:IndDir}

We first evaluate Eq. (\ref{CurMom}) directly.
In such a direct calculation, 
there is no need to specify the representation of the $\Gamma$-matrices.  
Note the relation, 
\begin{eqnarray}
e^{-ikx}{\cal H}^2e^{ikx}=-(\partial_j+ik_j)^2
+\phi_a^2-i\Gamma^j\Gamma^{d+a}\partial_j\phi_a .
\end{eqnarray}
Then, we have
\begin{eqnarray}
J^j(x) &=& \int\frac{{\rm d}^dk}{(2\pi)^d}{\rm tr}\,\Gamma^{2d+1}\Gamma^j(i\Gamma^{d+a}\phi_a)
\frac{1}{k_j^2+\phi_a^2+m^2-i\Gamma^j\Gamma^{d+a}\partial_j\phi_a}
\nonumber\\
&=& \int\frac{{\rm d}^dk}{(2\pi)^d}{\rm tr}\,\Gamma^{2d+1}\Gamma^j(i\Gamma^{d+a}\phi_a)
\sum_{l=0}^\infty\frac{(i\Gamma^j\Gamma^{d+a}\partial_j\phi_a)^l}{(k_j^2+\phi_a^2+m^2)^{l+1}}.
\label{CurExp}
\end{eqnarray}
In Eq. (\ref{CurExp}), 
the series expansion is quite useful, if we note that $\partial_j\phi_a=O(r^{-1})$
as well as ${\rm d} S_j\propto r^{d-1}{\rm d}^{d-1}\Omega$,
implying that terms with $l\ge d$ vanish when integrated over 
the surface of ${\rm S}^{d-1}$ in Eq. (\ref{IndThe}).
Note also that $\Gamma^{2d+1}$ appears in the above equation only once, and hence 
the terms with $l< d-1$ vanish if one takes the trace. It follows that 
only the $l=d-1$ term gives a finite contribution to the index, which is indeed well-defined.
At this point, one can clearly see that
the Pauli-Villars regulator for the current density, as expressed in Eq. (\ref{AxiVecCur}),
was indeed not necessary in the present case.
We reach
\begin{eqnarray}
J^j(x) &=& \int\frac{{\rm d}^dk}{(2\pi)^d}{\rm tr}\,\Gamma^{2d+1}\Gamma^j(i\Gamma^{d+a}\phi_a)
\frac{(i\Gamma^j\Gamma^{d+a}\partial_j\phi_a)^{d-1}}{(k_j^2+\phi_a^2+m^2)^{d}}
+O(r^{-d})
\nonumber \\
&=&
i^d\left[{\rm tr}\,\Gamma^{2d+1}\Gamma^j\Gamma^{d+a}
\Gamma^{j_1}\Gamma^{d+a_1}\Gamma^{j_2}\Gamma^{d+a_2}\cdots
\cdots\Gamma^{j_{d-1}}\Gamma^{d+a_{d-1}}\right]
\nonumber \\
&\times&
\phi_a\partial_{j_1}\phi_{a_1}\partial_{j_2}\phi_{a_2}\cdots
\partial_{j_{d-1}}\phi_{a_{d-1}}
\int\frac{{\rm d}^dk}{(2\pi)^d}\frac{1}{(k_j^2+\phi_a^2+m^2)^{d}}
+O(r^{-d}).
\end{eqnarray}
By the use of 
${\rm tr}\,\Gamma^{2d+1}\Gamma^{\mu_1}\Gamma^{\mu_2}\cdots\Gamma^{\mu_{2d}}
=(2i)^d\epsilon^{\mu_1\mu_2\cdots\mu_{2d}}$ for $\mu_j,\nu_j=1,\cdots,2d$, and
\begin{eqnarray}
\int\frac{{\rm d}^dk}{(2\pi)^d}\frac{1}{(k^2+h)^\alpha}
=\frac{\Gamma(\alpha-\frac{d}{2})}{(4\pi)^{\frac{d}{2}}\Gamma(\alpha)h^{\alpha-\frac{d}{2}}},
\end{eqnarray}
we have
\begin{eqnarray}
J^j(x) =
\frac{i^d(2i)^d\epsilon_d\Gamma(\frac{d}{2})}
{(4\pi)^{\frac{d}{2}}\Gamma(d)(\phi_a^2+m^2)^{\frac{d}{2}}}
\epsilon^{jj_1j_2\cdots j_{d-1}}\epsilon^{aa_1a_2\cdots a_{d-1}}
\phi_a\partial_{j_1}\phi_{a_1}\partial_{j_2}\phi_{a_2}\cdots
\partial_{j_{d-1}}\phi_{a_{d-1}} ,
\end{eqnarray}
where $\epsilon_d\equiv (-)^{d(d-1)/2}$, which is due to the rearrangement of the arguments
in the totally antisymmetric tensor,
$\epsilon^{j(d+a)j_1(d+a_1)\cdots(d+a_{d-1})j_{d-1}}
=\epsilon_d\epsilon^{jj_1\cdots j_{d-1}(d+a)(d+a_1)\cdots(d+a_{d-1})}$.
Let us assume that $\phi_a^2\rightarrow|\phi|^2\equiv\mbox{ const.}$ 
at $|x|\rightarrow\infty$. Then, 
\begin{eqnarray}
{\rm ind}\,{\cal H} &=&
\frac{(-)^{d}\epsilon_d\Gamma(\frac{d}{2})}{2\pi^{\frac{d}{2}}(d-1)!}
\int_{|x|\rightarrow\infty}
\epsilon^{jj_1j_2\cdots j_{d-1}}\epsilon^{aa_1a_2\cdots a_{d-1}}
\hat\phi_a\partial_{j_1}\hat\phi_{a_1}\partial_{j_2}\hat\phi_{a_2}\cdots
\partial_{j_{d-1}}\hat\phi_{a_{d-1}}{\rm d} S_j
\nonumber\\
&=&
\frac{(-)^{d}\epsilon_d\Gamma(\frac{d}{2})}{2\pi^{\frac{d}{2}}(d-1)!}
\int_{|x|\rightarrow\infty}
\epsilon^{a_1a_2\cdots a_{d}}
\hat\phi_{a_1}{\rm d}\hat\phi_{a_2}{\rm d}\hat\phi_{a_3}\cdots
{\rm d}\hat\phi_{a_{d}},
\label{TopIndGen}
\end{eqnarray}
where $\hat\phi_a=\phi_a/|\phi|$, and hence, $\hat\phi_a^2=1$.

So far our calculation was generic,
not depending on a specific form of the order parameter,
as in Eq. (\ref{Hig}).
Eq. (\ref{TopIndGen}) is a general expression of topological index
for an arbitrary configuration of the order parameter.
For a particular case of the hedgehog configuration, Eq. (\ref{Hig}), 
the integral in Eq. (\ref{TopIndGen}) 
reduces simply to the surface area of ${\rm S}^{d-1}$ with a unit radius, 
which leads to
\begin{eqnarray}
{\rm ind}\,{\cal H}=(-)^{d}\epsilon_d .
\label{TopIndHeg}
\end{eqnarray}
When $\Delta(\infty)>0$, 
this coincides exactly with
the analytical index in Eq. (\ref{AnaIndRes}).

\subsection{Relating the topological index to a transition function}
\label{s:IndTra}

In the last subsection we have directly evaluated Eq. (\ref{CurMom})
and obtained a couple of expressions for the topological index;
a general formula, Eq. (\ref{TopIndGen}) and
a concrete expression for the hedgehog configuration, Eq. (\ref{TopIndHeg}).
These expressions will be compared with the Chern number due to the Berry connection
in the next section \ref{s:CheDir}.
Here, to see a more direct relationship between these two,
we derive an alternative expression of the topological index.
This will correspond to 
the Chern number in section \ref{s:CheTra} expressed by the transition function.
To this end, let us introduce another specific representation of 
the $\Gamma$-matrices than the one we used for the evaluation of the analytical indices.
The representation we use here
makes chiral symmetry of the Hamiltonian manifest.

Let $\gamma^\mu$ be the $\gamma$-matrices in $2d-1$ dimensions given in 
Appendix \ref{s:Gam}, 
and let $\Gamma^\mu$ be the $\Gamma$-matrices in $2d+1$ dimensions which are
constructed by $\gamma^\mu$ such that
\begin{eqnarray}
\Gamma^\mu &=& \gamma^\mu\otimes\sigma^1, \quad(\mu=1,\cdots,2d-1),
\nonumber\\
\Gamma^{2d} &=& 1\otimes\sigma^2,
\nonumber\\
\Gamma^{2d+1}&=& 1\otimes\sigma^3.
\label{RepGamChi}
\end{eqnarray}
Regardless of $d$, the Hamiltonian turns out to be
\begin{eqnarray}
{\cal H}=
\left(\begin{array}{cc}&{\cal D}\\{\cal D}^\dagger&\end{array}\right) ,
\end{eqnarray}
where
\begin{eqnarray}
{\cal D}=-i\gamma^j\partial_j+\gamma^{d+a}\phi_a-i\phi_{d}.
\end{eqnarray}
We also introduce a matrix ${\cal Q}$ by 
\begin{eqnarray}
e^{-ikx}{\cal H}e^{ikx}={\cal Q}-i\Gamma\cdot\partial, 
\end{eqnarray}
where $\Gamma\cdot\partial=\Gamma^j\partial_j$. 
It is explicitly given by
\begin{eqnarray}
  {\cal Q}=\Gamma^jk_j+\Gamma^{d+a}\phi_a .
  \label{cQandDel}
\end{eqnarray}
For the $\Gamma$ matrices Eq. (\ref{RepGamChi}) 
we can write it as 
\begin{eqnarray}
{\cal Q}
=\left(\begin{array}{cc}&Q\\ Q^\dagger&\end{array}\right),
\quad Q=\gamma^jk_j+\gamma^{d+a}\phi_a-i\phi_{d}\bm1. 
\label{QandDel}
\end{eqnarray}
The matrix $Q$ plays the role of the transition function in Sec. \ref{s:CheTra}.
Indeed, by the use of 
$Q^\dagger Q=R_0^2\equiv \sum_{j=1}^dk_j^2+\sum_{a=1}^{d}\phi_a^2$, we can define 
the unitary matrix $\hat Q\equiv Q/R_0$.

Taking the limit $m\rightarrow0$, we see that 
the current (\ref{CurMom}) can be written as
\begin{eqnarray}
J^j(x)
\rightarrow
\int\frac{{\rm d}^dk}{(2\pi)^d}{\rm tr}\,\Gamma^{2d+1}
\Gamma^j\frac{1}{{\cal Q}-i\Gamma\cdot\partial}, 
\end{eqnarray}
where $j=1,\cdots,d$.
Note that $\Gamma\cdot\partial=\tilde\partial Q\cdot\partial$, 
where $\tilde\partial_j\equiv\partial_{k_j}$,
which is distinguished from $\partial_j\equiv\partial_{x_j}$.
Then, the expansion
\begin{eqnarray}
({\cal Q}-i\Gamma\cdot\partial)^{-1} &=& ({\cal Q}-i\Gamma\cdot\partial)
 \frac{1}{({\cal Q}-i\Gamma\cdot\partial)^2} 
\nonumber \\
&=& ({\cal Q}-i\Gamma\cdot\partial)
\frac{1}{R_0^2}\sum_{n=0}^\infty
\left((i\tilde\partial \ {\cal Q}\cdot\partial{\cal Q}+2ik\cdot\partial
+\partial^2)\frac{1}{R_0^2}\right)^n
\end{eqnarray}
yields the current 
\begin{eqnarray}
J^j(x)=
\int\frac{{\rm d}^dk}{(2\pi)^d}\frac{i^d}{R_0^{2d}}{\rm tr}\ 
\Gamma^{2d+1}\tilde\partial_j{\cal Q}{\cal Q}
(\tilde\partial{\cal Q}\cdot\partial{\cal Q})^{d-1}
+\mathrm{O}(|x|^{-d}) .
\label{Jjx}
\end{eqnarray}

Inserting this into the r.h.s. of Eq. (\ref{IndThe}), we 
can express the index of ${\cal H}$ as an integral over 
the $2d-1$ dimensional space, the $d-1$ dimensional sphere 
at the spatial infinities and the $d$ dimensional momentum 
space. The expression thus obtained, however, is not 
manifestly topologically invariant as it stands. 
It is possible to write $\mathrm{ind}\,{\cal H}$ 
as an integral of $(2d-1)$-form as 
\begin{eqnarray}
{\rm ind}\,{\cal H}=\frac{2\epsilon_d i^dd!}{(2\pi)^d(2d)!}
\int
{\rm tr}\,(\hat Q^\dagger{\rm d}\hat Q)^{2d-1}, 
\label{IndTra}
\end{eqnarray}
where $\hat Q = Q/R_0$ is the transition function and 
$\mathrm{d}=\mathrm{d}k_j\tilde\partial_j +\mathrm{d}x_j\partial_j$ 
is the exterior derivative with
respect to $x$ and $k$. The topological invariance 
now becomes manifest since the $(2d-1)$-form in the integral 
changes only by an exact form under any continuous deformation 
of the transition function. For reference, we will give 
a derivation of Eq. (\ref{IndTra}) in Appendix \ref{s:appb}.

\section{Topological indices vs. topological invariants (Berry curvature)}
\label{s:Che}

How (topological) indices are related to the topological invariants, 
i.e., Chern numbers,
expressed in terms of the Berry curvature?
In Refs. \cite{TeoKan10,TeoKane:10b},
a Chern number, analogous to the one in Refs.
\cite{TKNN:82, Kohmoto:85},
but characterizing the protected zero-modes,
has been proposed.
For bulk topological insulators and superconductors, 
Chern numbers given in terms of a Berry curvature 
in $\bm k$-space,
are used
as a topological quantum number of the ground state.
Here, the one proposed in Refs. \cite{TeoKan10,TeoKane:10b}
involves the Berry connection 
in both $\bm k$- and $\bm x$-space.

Let $H$ be the corresponding Hamiltonian proposed in Refs. \cite{TeoKan10,TeoKane:10b},
\begin{eqnarray}
H=\Gamma^jk_j+\Gamma^{d+a}\phi_a+\lambda\Gamma^{2d+1}\equiv\Gamma^\mu X_\mu ,
\end{eqnarray} 
where $\mu=1,2,\cdots,2d+1$, and $X_\mu=X_\mu(k_j,x_j,\lambda)$.
We assume that this model is defined at $|x|\rightarrow\infty$,
where $\phi_a^2=|\phi|^2=\mathrm{const}$, so that
it includes $d$ parameters $\{k_j\}$, $d-1$ parameters $\{x_j\}$
(these should be denoted by, say, $\theta_j$, the coordinates of $S^{d-1}$, 
but for simplicity, we use $x_j$), 
and one parameter $\lambda$, i.e, in total $2d$ parameters
$\{k_j,x_j,\lambda\}$.
Then, we see that $H^2=k_j^2+|\phi|^2+\lambda^2\equiv R^2$, 
which implies that the eigenvalues of this Hamiltonian are $\pm R$.
Let $\psi_j$ ($j=1,2,\cdots,2^{d-1}$) be a set of orthonormal 
eigenstates with eigenvalue $-R$, 
\begin{eqnarray}
H\psi_j=-R\psi_j .
\end{eqnarray}
Then, the Berry connection one-form and curvature two-form are, respectively, 
defined by 
\begin{eqnarray}
&
A=\psi^\dagger {\rm d}\psi,
\label{BerCon}\\ 
&
F={\rm d} A+A^2 ,
\label{FieStr}
\end{eqnarray} 
where $\psi=(\psi_1,\psi_2,\cdots,\psi_{2^{d-1}})$ is a 
$2^{d}\times2^{d-1}$ matrix and 
${\rm d}={\rm d} k_j \partial_{k_j}+{\rm d} x_j\partial_{x_j}+{\rm d}\lambda\partial_\lambda$.
It should be noted that the projection operator \cite{Hatsugai10} 
to the negative eigenstates can be written as
\begin{eqnarray}
P &=& \psi\psi^\dagger
\label{ProOpe1}\\ 
&=& \frac{1}{2}\left(1-\frac{H}{R}\right) .
\end{eqnarray}
Then, Eq. (\ref{ProOpe1}) leads to
$P({\rm d} P)^{2d}P=\psi F^d\psi^\dagger$, which immediately gives 
\begin{eqnarray}
{\rm tr}\ F^d={\rm tr}\ P({\rm d} P)^{2d} .
\end{eqnarray}
The $d$-th Chern number for $H$ is defined by
\begin{eqnarray}
c_d(H)=\frac{1}{d!}\left(\frac{i}{2\pi}\right)^d\int {\rm tr}\ F^d .
\label{CheNumBer}
\end{eqnarray} 
This is a topological invariant, to be sure, but
identifying it as  
the index is not more than a speculation.
As mentioned earlier,
Eq. (\ref{CheNumBer}) 
is expressed in terms of a Berry curvature 
which involves, in addition to $k_j$,
the coordinates $x_j$ surrounding the defect
as parameters for the Berry connection and curvature.
This suggests that the Chern number (\ref{CheNumBer}) characterizes 
the zero-energy states peculiar to the defect.
Though in $d=2$ and $3$, the topological index and the Chern number 
are shown to be equivalent,\cite{FukuiFujiwara:10} 
its direct proof is still missing.
Below we establish an explicit one-to-one correspondence between the two
quantities,
which is valid in arbitrary dimensions.
The two quantities are indeed shown to be equivalent, but
possibly differ by a sign factor depending on the spatial dimension.

\subsection{Direct calculation of the Chern number}
\label{s:CheDir}

One-form of the projection operator is
\begin{eqnarray}
&
{\rm d} P=-\frac{1}{2}\Gamma^\mu{\rm d}{\hat X}_\mu,
\end{eqnarray}
where $\hat X_\mu=X_\mu/R$.
Then, we have
\begin{eqnarray}
{\rm tr}\ F^d &=& -\frac{1}{2^{2d+1}}{\rm tr}\ \Gamma^{\mu}{\hat X}_{\mu}
\left(\Gamma^{\nu}{\rm d}{\hat X}_{\nu}\right)^{2d}
\nonumber\\
&=& -\frac{1}{2^{2d+1}}(2i)^d\epsilon^{\mu_1\mu_2\cdots\mu_{2d+1}}\frac{1}{R^{2d+1}}
X_{\mu_1}{\rm d} X_{\mu_2}\cdots{\rm d} X_{\mu_{2d+1}}.
\end{eqnarray}
Now let us divide $\{X_\mu\}$ into $\{\phi_a\}$ and $\{k_j\}$ 
(we regard $\lambda$ as the $(d+1)$-th momentum $k_{d+1}$). 
Here, note that the first $X_\mu$ in the above equation 
cannot be a momentum, since otherwise it vanishes 
after the integration over the momentum. So that
\begin{eqnarray}
\epsilon^{\mu_1\mu_2\cdots\mu_{2d+1}}
X_{\mu_1}{\rm d} X_{\mu_2}\cdots{\rm d} X_{\mu_{2d+1}}
&=& (-)^d\epsilon^{a_1a_2\cdots a_d}\epsilon^{j_1j_2\cdots j_{d+1}}
 ~_{2d}\mbox{C}_{d+1}
\phi_{a_1}{\rm d}\phi_{a_2}\cdots{\rm d}\phi_{a_d}
{\rm d} k_{j_1} {\rm d} k_{j_2}\cdots{\rm d} k_{j_{d+1}}
\nonumber\\
&=& \frac{(2d)!}{(d-1)!}(-)^d\epsilon^{a_1a_2\cdots a_d}
\phi_{a_1}{\rm d}\phi_{a_2}\cdots{\rm d}\phi_{a_d}
{\rm d}^{d+1}k .
\end{eqnarray}
Therefore, we have
\begin{eqnarray}
c_d(H) &=& \frac{(-)^{d-1}(-)^d}{d!(2\pi)^d2^{d+1}}\frac{(2d)!}{(d-1)!}
\int\epsilon^{a_1a_2\cdots a_d}\phi_{a_1}{\rm d}\phi_{a_2}\cdots{\rm d}\phi_{a_d}
\int\frac{{\rm d}^{d+1}k}{(k^2+|\phi|^2)^{d+\frac{1}{2}}}
\nonumber\\
&=& -\frac{\Gamma(\frac{d}{2})}{2\pi^{\frac{d}{2}}(d-1)!}
\int\epsilon^{a_1a_2\cdots a_d}\hat\phi_{a_1}{\rm d}\hat\phi_{a_2}\cdots{\rm d}\hat\phi_{a_d},
\end{eqnarray}
where $\hat\phi_a=\phi_a/|\phi|$.
It thus turns out that
\begin{eqnarray}
{\rm ind}\,{\cal H}=(-)^{d+1+d(d-1)/2}c_d(H) .
\label{IndAndBer}
\end{eqnarray}
This establishes an explicit one-to-one correspondence between
the topological index and the Chern number, 
which is valid for a generic configuration of order parameters. 
The two quantities are indeed equivalent but their relative sign shows a systematic
alternation as a function of $d$.
For $d=3$ this extra sign factor is typically $-1$.

\subsection{The transition function}
\label{s:CheTra}

Here, we give another derivation of  Eq. (\ref{IndAndBer}), i.e.,
by the use of the transition function.
The transition function, which will be denoted as $\hat{Q}$,
includes all the topological information of the Chern number.
The calculations presented in this subsection correspond to
those of Sec. \ref{s:IndTra}.
 
The $\Gamma$-matrices (\ref{RepGamChi}) lead to the Hamiltonian 
\begin{eqnarray}
H=\left(
\begin{array}{cc}
\lambda\bm1&Q\\Q^\dagger&-\lambda\bm1
\end{array}
\right) ,
\end{eqnarray}
where $Q$ is defined in Eq. (\ref{QandDel}).
Then, $Q^\dagger Q+\lambda^2\bm1=QQ^\dagger+\lambda^2\bm1=R^2\bm1$.
By the use of some reference states, we can fix the gauge of the wave functions such that
\begin{eqnarray}
\psi=P\phi N^{-1},
\end{eqnarray}
where $N^2=\phi^\dagger P\phi$ is a normalization matrix. 
Let us introduce two kinds of the reference states
\begin{eqnarray}
\phi_+ = \left(\begin{array}{c}0\\1\end{array}\right),\quad
\phi_- = \left(\begin{array}{c}-1\\0\end{array}\right) .
\end{eqnarray} 
Then, the normalization matrix is given by
\begin{eqnarray}
N_\pm=\sqrt{\frac{R\pm\lambda}{2R}}\bm1
\end{eqnarray}
and corresponding gauge-fixed wave function is given by
\begin{eqnarray}
\psi_+ &=& \frac{1}{\sqrt{2R(R+\lambda)}}
\left(\begin{array}{c}-Q\\(R+\lambda)\bm1\end{array}\right) ,
\nonumber\\
\psi_- &=& \frac{1}{\sqrt{2R(R-\lambda)}}
\left(\begin{array}{c}-(R-\lambda)\bm1\\Q^\dagger\end{array}\right).
\end{eqnarray}
The implication of the subscripts $\pm$ would be clear;
$\psi_\pm$ have singularities, respectively, in the south and north poles
on S$^{2d}$,
associated with the Dirac string.
It turns out that the transition function $g$ at $\lambda=0$ defined by
$\psi_+=\psi_-g$ is given by 
\begin{eqnarray}
g=\psi_-^\dagger\psi_+=\frac{Q}{R}\equiv \hat Q ,
\end{eqnarray}
where $\hat Q^\dagger \hat Q=\bm1$.
Using ${\rm tr}\ F^d={\rm d}\omega_{2d-1}(A,F)$, where $\omega_{2d-1}$ denotes the 
$2d-1$ th Chern-Simons form, and
$\omega_{2d-1}(A+V,F)-\omega_{2d-1}(A,F)=\omega_{2d-1}(V,0)+{\rm d}\alpha_{2d-2}$
with $V={\rm d} gg^{-1}$ and $\alpha_{2d-2}$ being a certain ($2d-2$)-form, we reach
\begin{eqnarray}
c_d=\frac{(-)^{d-1}i^d(d-1)!}{(2\pi)^d(2d-1)!}\int_{{\rm S}^{2d-1}}
{\rm tr}\,(\hat Q^\dagger{\rm d}\hat Q)^{2d-1} .
\end{eqnarray}
In view of Eq. (\ref{IndTra}), we have established the relation (\ref{IndAndBer}) 
again.

\section{Summary and discussions}
\label{s:Sum}

We have investigated
a Majorana zero-energy state in a topological superconductor
in arbitrary dimensions in order to reexamine the relation
between the topological index and
the Chern number 
associated with the Berry curvature.
We have first counted the analytical index by solving the BdG equation directly, and next 
calculated the topological index characterizing  the topological configuration of 
the order parameter. We have confirmed that they indeed coincide.
Based on these observations, 
we have finally calculated the Chern number also in arbitrary dimensions.
The obtained Chern number coincides with the topological index
apart from an extra sign factor depending on the dimension.
In this sense, the Chern number for systems with a topological defect is 
endowed with the role of the index for the protected zero-energy states.

The model we have studied in this paper 
belongs to class BDI with particle-hole, chiral, and resultant
time-reversal symmetries. 
Among them, chiral symmetry has played a  crucial role in the index theorem,
since the index is defined as eigenvalues of the chirality operator.
Note that the BDI class with a point defect 
(in the notation of Ref. \cite{TeoKane:10b},
$\delta = d-D =1$)
is classified by a $\mathbb Z$-type topological number,
which gives physically the number of protected zero-energy states.
A more generic class of similar character;
possessing particle-hole symmetry with non-trivial topological defects
is class D, which on the contrary, does not preserve chiral symmetry.
In class D protected zero-energy states are characterized by a
$\mathbb Z_2$-type topological number, i.e.,
the choice is simply, whether there exists one of such, or none.
This reduction of symmetry \cite{RSFL10} from class BDI to D is analogous to
the relation between IQHE (class A) and $\mathbb Z_2$ (quantum spin Hall)
insulators (class AII) in two spatial dimensions ($d=2$, 2D).

According to the classification by Teo and Kane \cite{TeoKane:10b},
there are other classes 
with various kinds of topological defects which allow protected zero modes.
It may be quite interesting to explore the relationship between the topological invariants
associated with the Berry connection and curvature, which may be valid, in a sense, in the 
adiabatic approximation, and index-like quantities directly reflecting the protected zero modes,
which are free from any approximations.

\acknowledgments
This work was supported in part by Grant-in-Aid for Scientific Research from JSPS
(No. 21540378) and by 
the ``Topological Quantum Phenomena'' 
Grant-in Aid for Scientific Research 
on Innovative Areas from the Ministry of Education, Culture, Sports, Science and Technology
(MEXT) of Japan (Nos. 23103502, 23103511).

\appendix

\section{$\gamma$-matrices}
\label{s:Gam}

In this appendix, we summarize the representation of the $\gamma$-matrices
useful to obtain the zero-energy wave functions in Sec. \ref{s:Zer}.
We first define the conventional $\gamma$-matrices in $2n+1$ dimensions such that
\begin{eqnarray}
\gamma^{2k-1} &=& \underbrace{1\otimes\cdots\otimes1}_{k-1}\otimes\,
\sigma^1\otimes\underbrace{\sigma^3\otimes\cdots\otimes\sigma^3}_{n-k}
\nonumber\\
\gamma^{2k} &=& \underbrace{1\otimes\cdots\otimes1}_{k-1}\otimes\,
\sigma^2\otimes\underbrace{\sigma^3\otimes\cdots\otimes\sigma^3}_{n-k}
\end{eqnarray}
where $k=1,\cdots,n$, and
\begin{eqnarray}
\gamma^{2n+1}
= \underbrace{\sigma^3\otimes\cdots\otimes\sigma^3}_n
= (-i)^n\gamma^1\cdots\gamma^{2n},
\end{eqnarray}
which obey $\{\gamma^\mu,\gamma^\nu\}=2\delta^{\mu\nu}$.
Define also,
\begin{eqnarray}
\Gamma=\gamma^1\gamma^3\cdots\gamma^{2n+1},
\label{gamma_cap}
\end{eqnarray}
and notice
\begin{eqnarray}
\Gamma^2=(-1)^{[(n+1)/2]}=(-1)^{n(n+1)/2}.
\end{eqnarray}
Then, we have
\begin{eqnarray}
(\gamma^\mu)^T=(\gamma^\mu)^*=(-)^{n}\Gamma\gamma^\mu\Gamma^{-1}, 
\label{ReaGam}
\end{eqnarray}
for $\mu=1,2,\cdots, 2n, 2n+1$.


\section{Derivation of Eq. (\ref{IndTra})}
\label{s:appb}

In the trace of Eq. (\ref{Jjx}), ${\cal Q}$, $\tilde\partial_j{\cal Q}$
and $\partial_j{\cal Q}$ effectively anti-commute one another 
because of the presence of $\Gamma^{2d+1}$. 
Noting this, we can put $J^j(x)$ as 
\begin{eqnarray}
  J^j(x) &=&\epsilon_d\int\frac{{\rm d}^dk}{(2\pi)^d}\frac{i^d}{R_0^{2d}}{\rm tr}\
[\Gamma^{2d+1}\tilde\partial_j{\cal Q}\tilde\partial_{j_2}{\cal Q}
\cdots\tilde\partial_{j_d}{\cal Q}
{\cal Q}\partial_{j_2}{\cal Q}\cdots\partial_{j_d}{\cal Q}] 
\nonumber \\
&=&\epsilon_d\int\frac{{\rm d}^dk}{(2\pi)^d}\frac{i^d}{R_0^{2d}}{\rm tr}\
[\Gamma^{2d+1}\tilde\partial_1{\cal Q}\tilde\partial_2{\cal Q}
\cdots\tilde\partial_d{\cal Q} 
\epsilon_{jj_2\cdots j_d}
{\cal Q}\partial_{j_2}{\cal Q}\cdots\partial_{j_d}{\cal Q}] \nonumber \\
&=&\frac{i^d\epsilon_d}{(2\pi)^dd!}\int
\frac{\epsilon_{jj_2\cdots j_d}}{R_0^{2d}}{\rm tr}\
[\Gamma^{2d+1}(\mathrm{d}_k{\cal Q})^d
{\cal Q}\partial_{j_2}{\cal Q}\cdots\partial_{j_d}{\cal Q}], 
\end{eqnarray}
where we have suppressed the nonleading terms for 
$|x|\rightarrow\infty$ and $\mathrm{d}_k=\mathrm{d}k_j\tilde\partial_j$ 
is the exterior derivative on the momentum space. Using this and the 
volume form on the $(d-1)$ dimensional sphere 
\begin{eqnarray}
  \mathrm{d}S^j=\frac{1}{(d-1)!}\epsilon^{jj_2\cdots j_d}
  \mathrm{d}x_{j_2}\cdots\mathrm{d}x_{j_d},
\end{eqnarray}
we can write Eq. (\ref{IndThe}) as  
\begin{eqnarray}
  \mathrm{ind}\,{\cal H} &=&\frac{i^d\epsilon_d}{2(2\pi)^dd!(d-1)!}\int
  \epsilon^{jj'_2\cdots j'_d}
  \mathrm{d}x_{j'_2}\cdots\mathrm{d}x_{j'_d} 
\int\frac{\epsilon_{jj_2\cdots j_d}}{R_0^{2d}}{\rm tr}\
[\Gamma^{2d+1}(\mathrm{d}_k{\cal Q})^d
{\cal Q}\partial_{j_2}{\cal Q}\cdots\partial_{j_d}{\cal Q}] 
\nonumber \\
&=&\frac{(-i)^d\epsilon_d}{2(2\pi)^dd!}\int\frac{1}{R_0^{2d}}
{\rm tr}\ [\Gamma^{2d+1}{\cal Q}(\mathrm{d}_k{\cal Q})^d
(\mathrm{d}_x{\cal Q})^{d-1}] 
\nonumber \\
&=&\frac{(-i)^d\epsilon_d}{(2\pi)^d}\frac{d!}{(2d)!}
\int\frac{1}{R_0^{2d}}{\rm tr}[\Gamma^{2d+1}{\cal Q}(\mathrm{d}{\cal Q})^{2d-1}],
\label{apBi}
\end{eqnarray}
where we have introduced exterior derivatives by 
$\mathrm{d}_x=\mathrm{d}x_j\partial_j$ and 
$\mathrm{d}=\mathrm{d}_x+\mathrm{d}_k$. In the
representation Eq. (\ref{RepGamChi}) of $\Gamma$ matrices
and for ${\cal Q}$ given by Eq. (\ref{QandDel}), 
the integrand can be further simplified as 
\begin{eqnarray}
  \frac{1}{R_0^{2d}}{\rm tr}[\Gamma^{2d+1}{\cal Q}(\mathrm{d}{\cal Q})^{2d-1}]
&=& \frac{1}{R_0^{2d}}{\rm tr}[Q\mathrm{d}Q^\dagger
  (\mathrm{d}Q\mathrm{d}Q^\dagger)^{d-1}
  -Q^\dagger\mathrm{d}Q(\mathrm{d}Q^\dagger\mathrm{d}Q)^{d-1}]
\nonumber \\
&=& (-1)^d2{\rm tr}(\hat Q^\dagger\mathrm{d}\hat Q)^{2d-1}+\mathrm{d}(\cdots),
  \label{apBtr}
\end{eqnarray}
where $\mathrm{d}(\cdots)$ denotes total derivative terms and has 
no contribution to $\mathrm{ind}\,{\cal H}$. Eq. (\ref{IndTra}) 
immediately follows from Eqs. (\ref{apBi}) and (\ref{apBtr}).

\end{document}